\documentclass[pra,superscriptaddress,twocolumn]{revtex4}
\usepackage{blindtext}
\usepackage{amsfonts}
\usepackage{amsmath,amssymb}
\usepackage{url}
\usepackage{hyperref}
\hypersetup{
     colorlinks = true,
     linkcolor = blue,
     anchorcolor = blue,
     citecolor = blue,
     filecolor = blue,
     urlcolor = blue
}
\usepackage{graphicx}
\usepackage{xcolor}
\usepackage{subfigure}
\usepackage{bm}
\usepackage{epstopdf}
\begin{document}
\title{Relaxation of ferromagnetic domains in a disordered lattice in 2D}
\author{C. Madro\~nero}
\author{G.A. Dom\'{\i}nguez-Castro}
\author{L. A. Gonz\'alez-Garc\'{\i}a}
\author{R. Paredes}
\email{rosario@fisica.unam.mx}
\affiliation{Instituto de F\'{\i}sica, Universidad
Nacional Aut\'onoma de M\'exico, Apartado Postal 20-364, M\'exico D.
F. 01000, Mexico. }

\begin{abstract}
We investigate the relaxation process of ferromagnetic domains in 2D subjected to the influence of both, static disorder of variable strength and weak interactions. The domains are represented by a two species bosonic mixture of $^{87}$Rb ultracold atoms, such that initially each specie lies on left and right halves of a square lattice. The dynamics of the double domain is followed by describing the two-component superfluid, at mean field level, through the time dependent Gross-Pitaevskii coupled equations, considering values of the intra and inter-species interaction, reachable in current experimental setups, that guaranty miscibility of the components. A robust analysis for several values inter-species interaction leads us to conclude that the presence of structural disorder leads to slowdown the relaxation process of the initial ferromagnetic order. As shown by our numerical experiments, magnetization is maintained up to 60 percent of its initial value for the largest disorder amplitude.  
\end{abstract} 

\maketitle

\section{Introduction}
\label{intro}
As much of the many-body problems within the condensed matter field, magnetism and particularly the dynamics of microscopic spins lying in ultrathin films in definite regions of space, the so-called magnetic domains dynamics, remain until now as an open question \cite{Himpsel, Kim, Thomson}. The origin of this dynamics can be attributed to several factors, for instance, the presence of external drivings as magnetic or electric fields, the existence of spin-polarized currents inducing the transference of momentum to the domain wall \cite{Thiaville, Franke}, or the inner dynamics associated with both, the interactions between the microscopic constituents as well as the energetic landscape where the constituents move. In the present investigation, we concentrate on analyzing the dynamics of the magnetic domains arising from the last causes. In particular, we focus on the effects of energy disorder in preventing the motion of spin domains.  

\begin{figure}[h]
\begin{center}
\includegraphics[width=8cm, height=4cm]{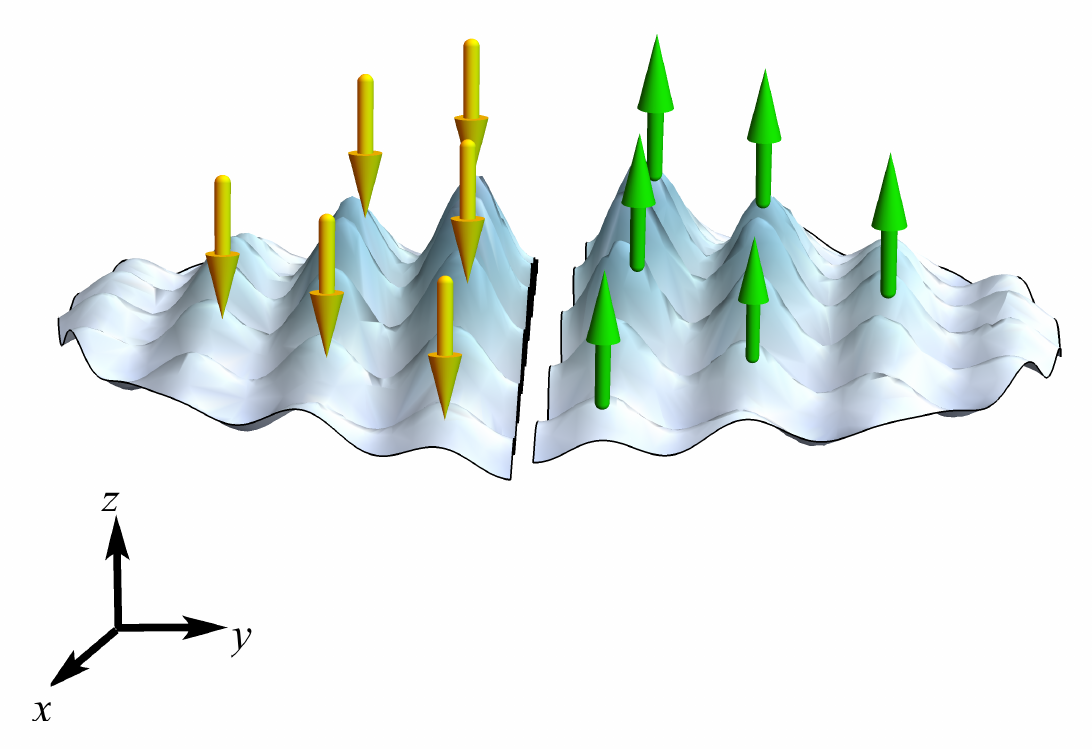}
\end{center}
\caption{Schematic form of the density profile prepared as the initial state. Left and right sides represent the superfluid density associated with the hyperfine components $\uparrow$ and $\downarrow$. Such profiles correspond to densities at zero disorder amplitude and a given value of the intra and inter-species interactions amplitudes $g_{\uparrow \uparrow}= g_{\downarrow \downarrow}$, and $g_{\uparrow \downarrow}$.}
\label{Figure1}
\end{figure}

Motivated by the notable control achieved with large conglomerates of atoms in its quantum degenerate state, and particularly the production of mixtures composed of either, Bose condensates in different hyperfine states \cite{Myatt}, or different atomic species \cite{Modugno, Thalhammer, Lercher, McCarron,Wang}, confined in particular geometries \cite{Hinds,Grimm,Lewenstein,Gross,LaRooji}, we propose here the design of an {\it ultracold atom device} to quantum simulate the decay of magnetization in magnetic domains in disordered square lattices in 2D. Our proposal is based on various experimental situations, previously performed with $^{87}$Rb atoms, intended to explore the many-body localization phenomenon \cite{Choi, IBloch}.  In particular, in \cite{Choi}, the initial state prepared is a Bose condensate composed of about one hundred atoms confined in a 2D square lattice in its Mott equilibrium state, and then allowed to evolve in a disordered potential under its own dynamics after suddenly changing an external parameter.
 Such a quantum quench protocol planned to track the effects of disorder on the atom flux moving across the 2D lattice, together with the possibility of spatially separating different hyperfine components, are the basis of our proposal to study the dynamics of the ferromagnetic domains, particularly its magnetization decay. As we describe below, in this work we shall consider a two-species Bose condensate as the analog of a double spin domain in which each hyperfine component lies in the halves of an inhomogeneous square lattice, thus setting initial configuration that will evolve in a disordered media (see Fig. \ref{Figure1}). This arrangement together with a recent study, performed at mean field level, where the effect of disorder is to induce the emergence of spatially localized densities, as a function of the disorder magnitude \cite{Gonzalez}, are our starting point to study the dynamics of the double spin domain. The idea here presented can be extended to other geometrical configurations, like those considered in \cite{Gonzalez} that can be useful for practical purposes as for instance the design of spin based magnetic protocols with complex configurations.
 
Here we present the results of an extensive set of numerical calculations performed, at the mean-field level through the coupled Gross-Pitaevskii (GP) equations, to describe the evolution in time of the hyperfine spin components spatially separated at $t=0$, and then allowed to evolve under the influence of non-correlated static disorder. Working within the superfluid regime, which is considering values of the intra-species interaction coupling for which the system is far from the Mott insulating phases (MI), we analyze the evolution of the initial state for different values of the ration among intra and inter-species interaction strengths.

This work is organized as follows. In section 2, we present the model that we use to describe the relaxation of the ferromagnetic domains under the influence of disorder. Furthermore, we briefly explain the construction of the initial state from which the evolution in time is followed. In section 3 we show the results of our numerical study about the relaxation process of the ferromagnetic domains, as a function of the disorder amplitude and different interaction strengths. Finally, in section 4, we summarize our findings.

\section{Model and initial state preparation}
\label{section2}

The model here proposed to study the persistence of magnetization in definite regions of space, is based, as described previously, on a series of experimental designs created with ultracold $^{87}$Rb atoms confined in 2D optical lattices, and their remarkable attribute of generating localized states as a result of both, disorder and two-body interactions. Here we concentrate on weakly interacting systems subjected to disorder. The system under study consists of a mixture of two hyperfine spin components, $|\uparrow \rangle= |F=1,m_F=-1\rangle$ and $|\downarrow \rangle=|F=2,m_F=-2\rangle$, lying in a 2D inhomogeneous square lattice, represented by $V_{ \mathrm{ext}}\left(\vec {r}\right)$. Within the mean-field formalism the wave functions $\Psi_{\uparrow,\downarrow}$ of the two species $|\uparrow \rangle$ and $|\downarrow \rangle$ obey the following effective coupled GP equations:
\small
\begin{eqnarray} 
i\hbar \frac { \partial \Psi _{\uparrow} (\vec {r},t)}{ \partial t } =\left[ H_0(\vec {r}) +  g_{\uparrow\uparrow}|\Psi_{\uparrow}|^{2} + g_{\uparrow\downarrow}|\Psi_{\downarrow}|^{2} \right]  \Psi_{\uparrow}(\vec{r} ,t)\cr
i\hbar \frac { \partial \Psi _{\downarrow} (\vec {r},t)}{ \partial t } =\left[ H_0(\vec {r})  +   g_{\downarrow\downarrow}|\Psi_{\downarrow}|^{2} + g_{\downarrow\uparrow}|\Psi_{\uparrow}|^{2} \right]  \Psi_{\downarrow}(\vec{r} ,t),
\label{coupledGP}
\end{eqnarray}
where $H_0(\vec {r})= -\frac { \hbar^ 2 }{2m} \nabla_{\perp}^ 2 +V_{ \mathrm{ext}}\left(\vec {r}\right)$ with $\nabla_\perp^2=\frac{\partial^2}{\partial x^2}+\frac{\partial^2}{\partial y^2}$ is the Laplacian operator in $2$D and $m$ the equal mass of the two spin components. The external potential in 2D has the following form:
\begin{eqnarray}
V_{ \mathrm{ext}}\left(\vec {r}\right)= \frac{1}{2} m \omega_r^2 r^2+V_{0}^\delta  \Bigg[  \sin^2 \left({\frac{\pi x}{a}}\right)+  \sin^2 \left({\frac{\pi y}{a}}\right) \Bigg],
\end{eqnarray}
\normalsize
being $\vec {r}= x \hat i +y \hat j$, $\omega_{r}$ the radial harmonic frequency, which is fixed to a common value used in current experiments $\omega_{r} = 2\pi\times 50$ Hz, $a$ is the lattice constant and $V_{0}^\delta= V_{0}(1+{\epsilon_\delta(x,y)})$, the potential depth at each point $(x,y)$. The function $\epsilon_{\delta} (x, y)$ represents a non-correlated disorder spanned across space and takes random values in the interval $\epsilon_{\delta} (x, y) \in [-\delta,\delta]$, being $\delta$ the disorder amplitude $\delta \in [0,1]$.
The random depth $V_{0}^\delta$ mimics the disordered environment introduced by speckle patterns \cite{Bouyer} and is scaled in units of the recoil energy $E_R= \frac{\hbar^2 k^2}{2m}$, with $k=\pi / a$. Thus, besides the contribution of the harmonic confinement, the potential depth at each point $(x,y)$ is the result of adding/subtracting a random number $\epsilon_{\delta}(x,y)$ to the amplitude of the potential defining the square lattice at zero disorder. Several previous studies have shown that the mean-field approximation describes the main effects of weakly interacting disordered systems \cite{Ray, Schulte, Adhikari, Kobayashi, Gonzalez}.

The values of the effective interaction couplings $g_{\sigma\sigma'}$ with $\sigma, \sigma' = \{ \uparrow, \downarrow \}$ are written in terms of the s-wave scattering length $a_{\sigma, \sigma'}$ as, $g_{\sigma\sigma'}= 4\pi N\hbar^{2}a_{\sigma\sigma'}/m$, being $N$ the number of particles in the condensate. We should point here that these interaction coefficients must be substituted by effective interaction couplings that take into account that the atom collision processes occur in 2D \cite{U2D, Salasnich, Mateo, Bao, Trallero, Zamora, U2D2}. The effective scattering length in the plane $x-y$ becomes $a_{\sigma\sigma'} \rightarrow a_{\sigma\sigma'}/\sqrt{2 \pi} l_z$, with $l_z=\sqrt{\hbar/m \omega_z}$, being $\omega_z$ a typical frequency of condensates confined in 2D \cite{Hadzibabic, Lung-Hung}. In typical experiments the values of the coupling constants $g_{\sigma\sigma'}$, can be varied via Feshbach resonances, and thus adjusted to have either, equal or different values of the intra and inter-species interactions, that is $g_{\uparrow \uparrow}= g_{\downarrow \downarrow}= g_{\uparrow \downarrow}$, or $g_{\uparrow \uparrow}=g_{\downarrow \downarrow} \neq g_{\uparrow \downarrow}$. In the present investigation we consider $g_{\uparrow \uparrow}= g_{\downarrow \downarrow}$ and $g_{\uparrow \downarrow}= g_{\downarrow \uparrow}$. As stated in \cite{Wang, Papp}, miscibility of the two component mixture is determined by the relation between $g_{\uparrow \uparrow}$, $g_{\downarrow \downarrow}$ and $g_{\uparrow \downarrow}$, as a matter of fact, the separation of the hyperfine components happens when the condition $g_{\uparrow \downarrow} > \sqrt{g_{\uparrow \uparrow} g_{\downarrow \downarrow}}$ is satisfied. In our analysis we shall consider values of the intra and inter-species interaction that guaranty miscibility of the hyperfine components.

\subsection{Initial ferromagnetic state}
To set the initial state from which we shall follow the dynamics to track the demagnetization process, we first solve the coupled equations (\ref{coupledGP}) for an optical lattice with lattice spacing $a = 532$ nm and a depth $V_{0}/E_R = 4$ without disorder, that is $\delta = 0$ and determine the stationary state. For this purpose free energy minimization is performed by means of imaginary time evolution  $\tau\rightarrow it$  \cite{Calculations,Dum,Zhang}. After this procedure, we manually remove the particles having spin component $\sigma =\uparrow$ from the left half layer, while particles with $\sigma=\downarrow$ from the right half layer, see Fig.\ref{Figure1}. This removal of particles mimics experimental procedures in which a digital mirror device is used to optically remove the particles at specific positions \cite{Choi}. Another route to achieve ferromagnetic domains is by means of a magnetic field \cite{Weld}. Notice that in our case similar ground state densities of different hyperfine states remain at each half in the 2D lattice. This pattern manually created, that is the two ferromagnetic domains, is our starting point to study its time evolution under the influence of static non-correlated disorder. We should note here that such an initial state is non-stationary, and consequently it evolves under their own dynamics. Interestingly, this kind of states from which a system evolves under its own dynamics are the so called quantum quenches created in the laboratory. Our particular interest is to investigate how the local magnetization of the ferromagnetic domains degrades when the weakly interacting 2D Bose mixture evolves in the absence of other external fields, except the one produced from the combination of a speckle pattern and the square lattice. It is important to mention that the effective coupling interaction coefficients must be rescaled for $t>0$ since half of the population is removed to have the magnetic domains, that is $N \rightarrow N/2$.

\section{Results: Demagnetization vs. disorder}
\label{section3}
With the purpose of establishing how magnetic domains demagnetize, we follow the time dynamics of an initial state prepared as described in the previous section, namely, spatially separated populations of the hyperfine components $\downarrow$ and $\uparrow$ lying in left and right sides of the 2D square lattice respectively. In Fig.\ref{Figure1} we show a schematic plot of the initial state. In our simulations we consider $N=300$ giving rise to interaction strengths $g_{\uparrow\uparrow}= g_{\downarrow\downarrow} = 10$, and lattices having $\sim 30 \times 30$ occupied sites. As stated in section \ref{section2}, the non-correlated disorder is introduced across the whole lattice through the function $\epsilon_{\delta}(x,y)$. To perform a reliable analysis of the physical quantities and have meaningful predictions, we take the average over an ensemble of 200 realizations for each value of the disorder amplitude $\delta$, and given values of the ratio among intra and inter-species interactions $g_{\uparrow \uparrow}/g_{\downarrow\uparrow}$. The purpose behind considering multiple realizations of disorder for a given value of the disorder strength, is to recreate the situation of experiments in which, typically, a single random realization could be not sufficient to represent the usual behavior of multiple scattering events due to uncorrelated disorder. We should notice that the way in which the disorder has been simulated, warrants that although the lattice symmetry is altered, the underlying structure is preserved, that is, the square geometry and the harmonic confinement prevail. Also, we must point out that the initial state depends on the particular values that the intra and inter-species interactions have.

The observables to be studied in our analysis are the magnetization in left and right sides $m_L$ and $m_R$ as a function of time. These quantities are defined in terms of the local magnetization $m(x,y;t) = \rho_{\uparrow}(x,y;t)-\rho_{\downarrow}(x,y;t)$, where $\rho_{\uparrow}(x,y;t)$ and $\rho_{\downarrow}(x,y;t)$ are the densities associated with the components $\uparrow$ and $\downarrow$ respectively. Thus, magnetization in left and right sides are,
\begin{eqnarray}
m_L=\int  \int_{\Omega_L} dx \ dy \> m(x,y;t)\cr
m_R=\int  \int_{\Omega_R} dx \ dy \> m(x,y;t),
\end{eqnarray}
where $\Omega_R$ and $\Omega_R$ are the left and right halves of the system, respectively. Because of the particular election of the initial state we have that $m_L(t=0) =-0.5$ and $m_R(t=0)=0.5$. For our analysis, besides the set of random realizations $\epsilon_{\delta}(x,y)$ for a given disorder magnitude, we shall consider three different values of ratio $g_{\uparrow \uparrow}/ g_{\downarrow\uparrow}$. We shall identify the dimensionless time in all of our calculations as $\tau= E_R t/\hbar$. The time dynamics was followed for a period of time such that at zero disorder and a given value of the coupling interactions the magnetization in left and right sides become null. As we shall see, the elapsed time during which the time dynamics is studied depends on the values of the inter- and intra-particle interaction strengths. That is, the relaxation of magnetization at zero disorder is different for each ratio $g_{\uparrow \downarrow}/g_{\uparrow \uparrow}$ considered. It is important to mention here that all of our numerical calculations were performed ensuring that changing $\tau \rightarrow -\tau$, at any temporal step along the time dynamics, allow us to recover the initial state. 
Furthermore, we found that the high precision used in our numerical simulations allowed us to observe a kind of macroscopic spin oscillation phenomenon, namely, that in the absence of disorder, the domains oscillate back and forth for extremely long evolution times.

\begin{figure}[h]
\begin{center}
\includegraphics[width=10.5cm, height=8cm]{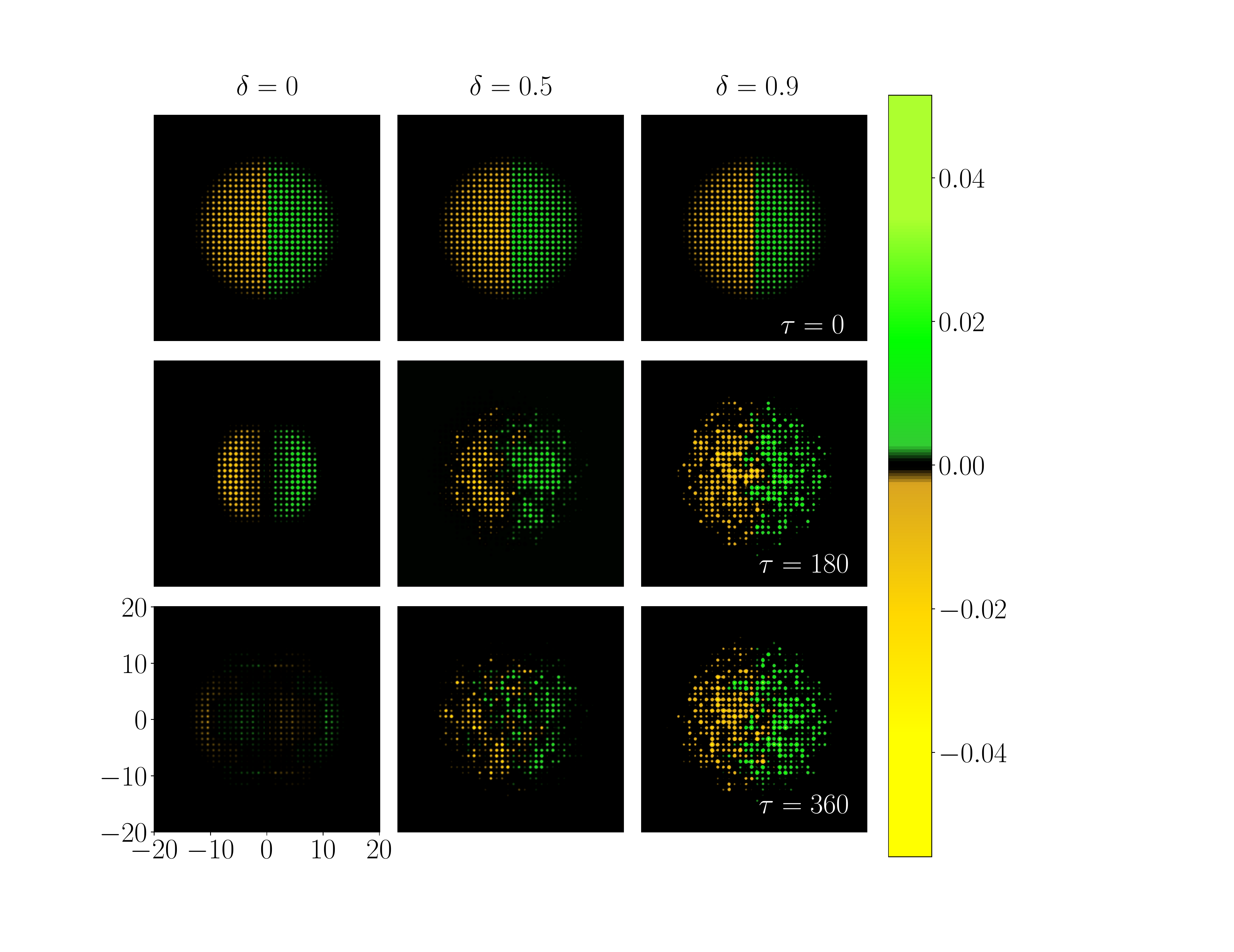}
\end{center}
\caption{Snapshots of the local magnetization in the square lattice for three different values of the disorder amplitude. Left, center and right columns correspond to three different values of disorder amplitude $\delta$ as indicated in the figure. Upper, middle and bottom rows are associated to $\tau=0$, $\tau=180$ and $\tau=360$ respectively. The ratio of the intra and inter-species interaction is, $g_{\uparrow \downarrow}= 0.9 g_{\uparrow \uparrow}$.}
\label{Fig_density}
\end{figure}

Since the prepared initial state is non-stationary, the hyperfine spin populations will evolve under the influence of both, disorder and interactions. Previous analysis of a single BEC component confined in a disordered square in 2D have shown that, in the weakly interacting regime, the net effect of the disorder is to localize the condensate density in bounded regions \cite{Gonzalez}. As a matter of fact, the size of those bounded regions become shorter and shorter as the amplitude of the disorder strength is increased. Therefore, what we expect in the case of the two component condensate is to have spatially localized densities of the condensate as the disorder magnitude grows, and thus preservation of magnetic domains. In Fig. \ref{Fig_density} we show snapshots of the local magnetization for three different values of the disorder amplitude, $\delta=0$, $\delta= 0.5$ and $\delta=0.9$ (left, center and right columns respectively), and three different times along the dynamics, $\tau=0$, $\tau=180$ and $\tau=360$. Each plot is a snapshot associated to a given realization of disorder $\epsilon_{\delta}(x,y)$ and fixed value of the intra and inter-species interaction ratio, $g_{\uparrow \downarrow}/g_{\uparrow \uparrow} = 0.9$. As one can see from this figure, at zero disorder amplitude, $ \uparrow$ and $\downarrow$ density configurations remain exactly opposite, while showing an asymmetric behavior for $\delta \neq 0$. We also observe how larger values of the disorder amplitude leads to a slow down the dynamics of the magnetic domains. That is, the initial state persists for larger times, thus showing a persistence of the magnetic domains during the time evolution. Fig. \ref{Fig_density} is representative of the magnetization behavior observed at different times as disorder is increased for the ensemble of disorder realizations. In the next paragraphs we outline the findings of our numerical experiments.

\begin{figure}[h]
\begin{center}
\includegraphics[width=7cm, height=7cm]{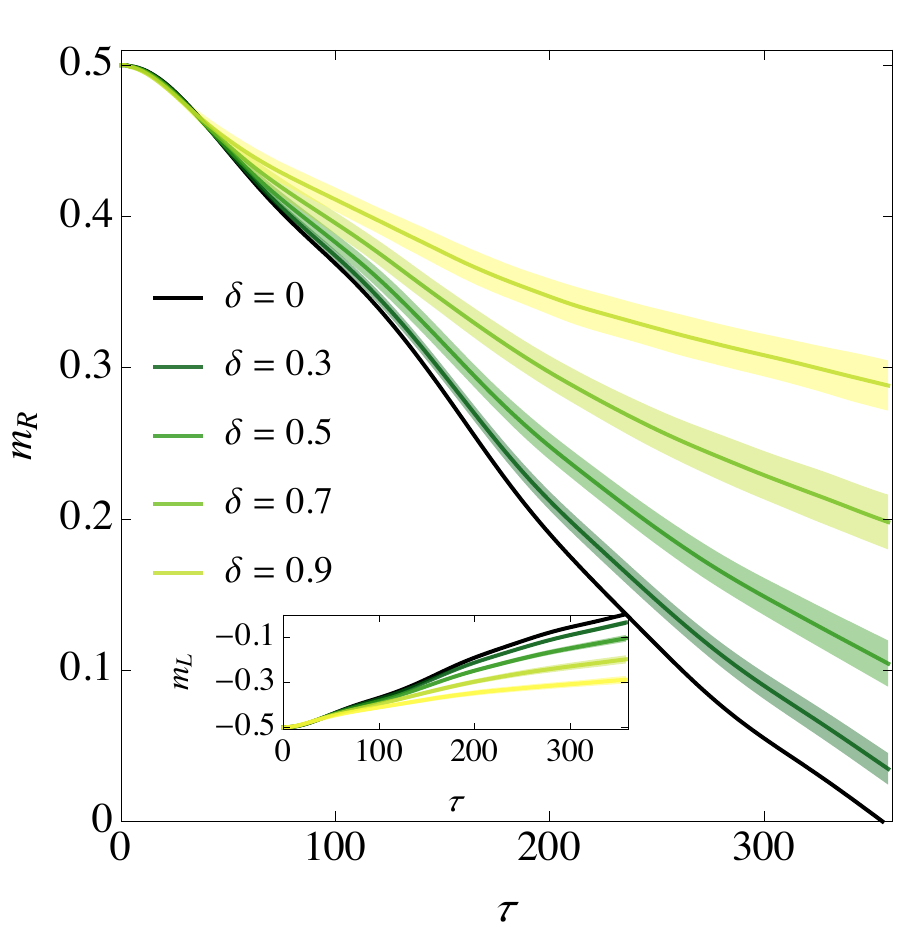}
\end{center}
\caption{Magnetization on the right side as a function of time for different values of disorder amplitude $\delta$, colors indicate the size of such a disorder. The ratio of the intra and inter-species interaction is $g_{\downarrow\uparrow}/g_{\uparrow \uparrow} = 0.9$. Magnetization in the left side is shown in the inset. Each point in these curves is the result of the average over 200 realizations of disorder for a given value $\delta$. The shadow area around each curve corresponds to the root mean square deviation.}
\label{Figure2}
\end{figure}

\begin{figure}[h]
\begin{center}
\includegraphics[width=7cm, height=7cm]{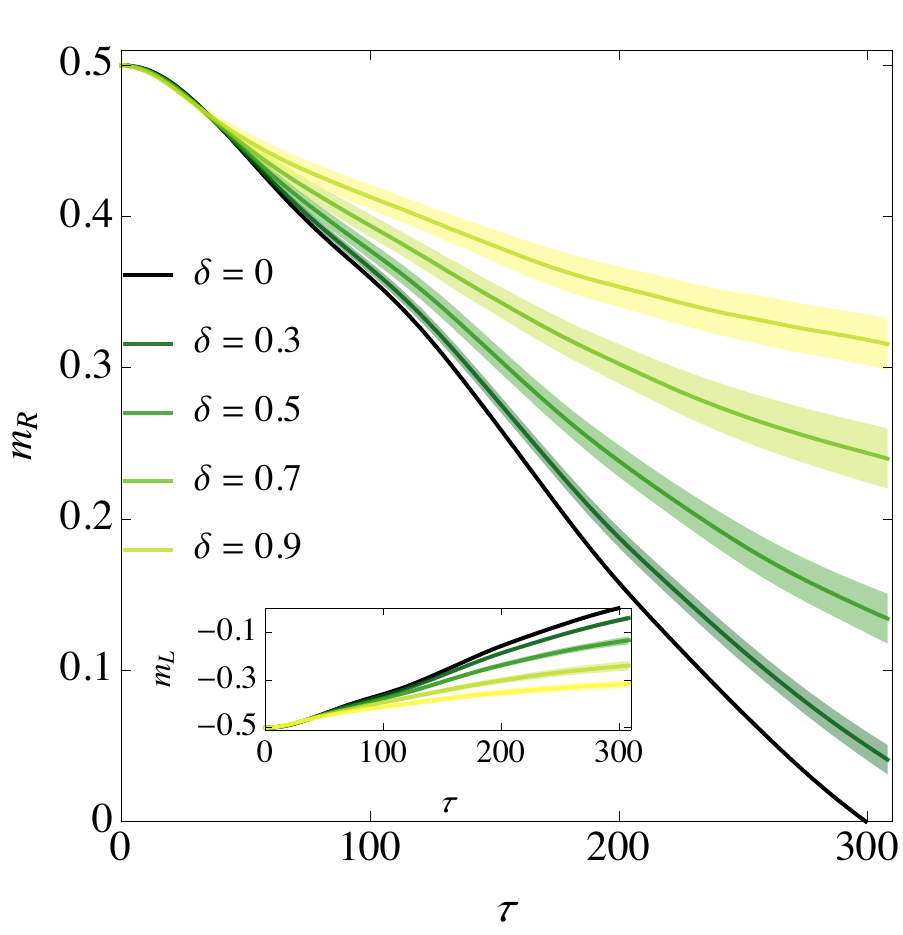}
\end{center}
\caption{Magnetization in the right side as a function of time for different values of disorder amplitude $\delta$, colors indicate the size of such a disorder. The ratio of the intra and inter-species interaction is $g_{\downarrow\uparrow}=0.8 g_{\uparrow \uparrow}$. Magnetization in the left side is shown in the inset. Each point in these curves is the result of the average over 200 realizations of disorder for a given value $\delta$. The shadow area around each curve corresponds to the root mean square deviation.}
\label{Figure3}
\end{figure}

\begin{figure}[h]
\begin{center}
\includegraphics[width=7cm, height=7cm]{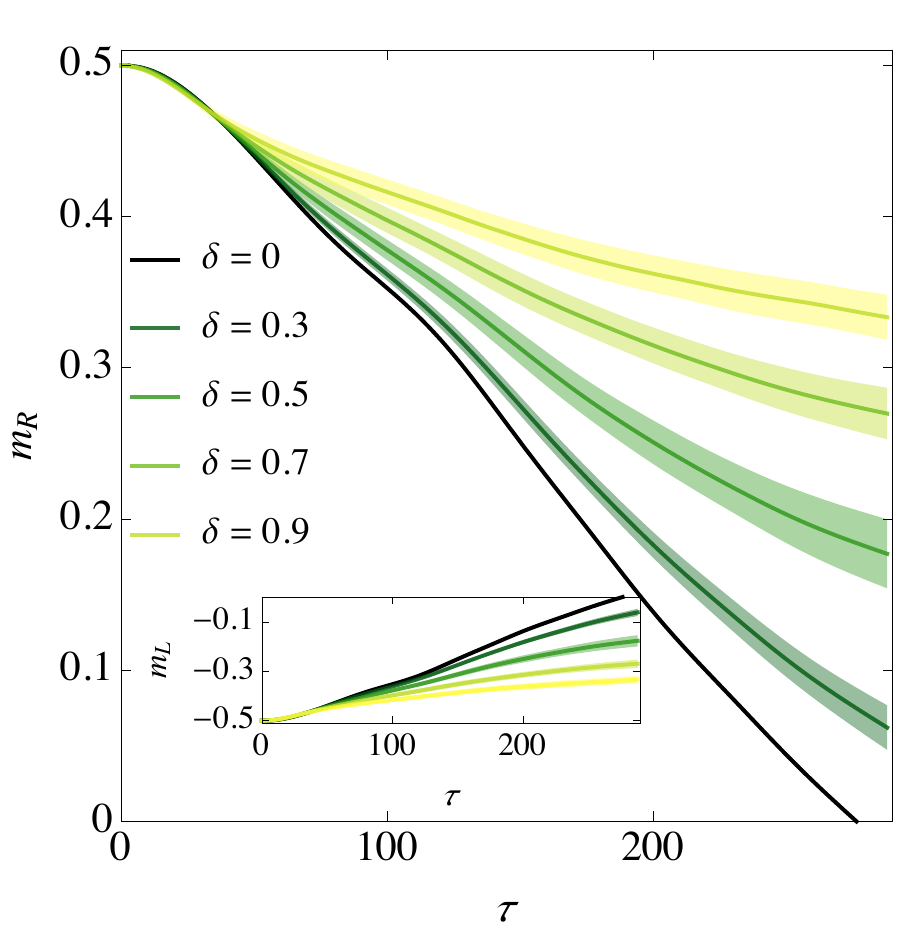}
\end{center}
\caption{Magnetization on the right side as a function of time for different values of disorder amplitude $\delta$, colors indicate the size of such a disorder. The ratio of the intra and inter-species interaction is $g_{\downarrow\uparrow}=0.7 g_{\uparrow \uparrow}$. Magnetization in the left side is shown in the inset. Each point in these curves is the result of the average over 200 realizations of disorder for a given value $\delta$. The shadow area around each curve corresponds to the root mean square deviation.}
\label{Figure4}
\end{figure}

As described above, to determine the influence of both, disorder and interactions, we track the evolution of the magnetization for an ensemble of realizations for a given value of the disorder amplitude. We restate that the values of the intra and inter-species interaction considered in our study are such that the miscibility of hyperfine components can occur. In Figures \ref{Figure2}, \ref{Figure3} and \ref{Figure4} we summarize the results of our analysis. Each plot correspond to the average of the magnetization on the right (main plot) and left (inset) sides as a function of time for different values of disorder amplitude $\delta$. The specific values of $\delta$ are indicated in the figures. The values of the intra and inter-species interaction of figures \ref{Figure2}, \ref{Figure3} and \ref{Figure4} are $g_{\downarrow\uparrow}=0.9 g_{\uparrow \uparrow}$, $g_{\downarrow\uparrow}=0.8 g_{\uparrow \uparrow}$ and $g_{\downarrow\uparrow}=0.7 g_{\uparrow \uparrow}$ respectively. Different curves in each plot are the average over the realizations for a given values of $\delta$, being the shadow area around each curve associated with the root mean square deviation. From these figures, one can observe that for short times $\tau \lesssim 50$, the general relaxation behavior of the magnetization takes quite similar values, independent of the disorder strength. However, for later times $\tau \gtrsim 50$, each magnetization curve departs from each other, thus revealing the effects of the disordered media. Furthermore, as the disorder amplitude is increased, the ferromagnetic order in each domain is preserved against the relaxation process. This slow relaxation dynamics induced by disorder enhances memory-like effects in the coupled magnetic domains. In particular, for the largest value of the disorder amplitude considered, that is $\delta = 1$, the value of the magnetization in left and right sides remains unaltered around 60\%. One can also notice two main outcomes associated with the value of the ratio $g_{\downarrow \uparrow}/g_{\uparrow \uparrow}$. The first is that at zero disorder the relaxation time is decreased as the ratio $g_{\downarrow \uparrow}/g_{\uparrow \uparrow}$ is diminished. The second one is that as this ratio is increased, the magnetization becomes progressively worse with respect to its initial value.

\begin{figure}[h]
\begin{center}
\includegraphics[width=8cm, height=5.5cm]{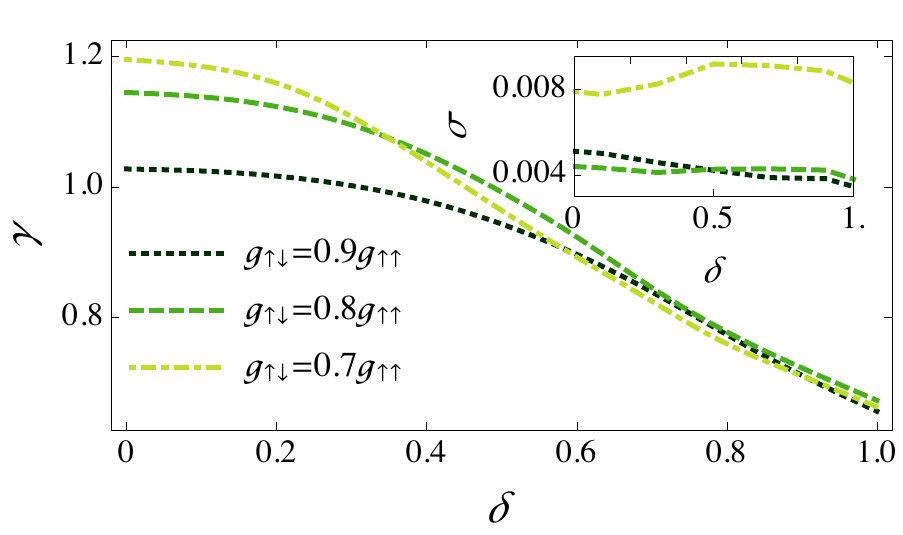}
\end{center}
\caption{Power-law fit $m_{R}(\tau)\propto b(\delta)\tau^{\gamma (\delta)}$ for the magnetization on the right side as a function of $\delta$ for three different values of the ratio of the intra and inter-species interaction.}
\label{Figure5}
\end{figure}

The time dependence of the left and right magnetizations are well described by a power-law ansatz $m_R(\tau)\propto b(\delta)\tau^{\gamma (\delta)}$. We fit the curves of figures \ref{Figure2}, \ref{Figure3} and \ref{Figure4} with such a power-law ansatz at intermediate times scales, where one neglects the transient behavior at short times. We should notice that the coefficients $\gamma$ and $b$ also depend upon the ratio among $g_{\downarrow\uparrow}$ and $g_{\uparrow \uparrow}$. In Figure \ref{Figure5} we plot the behavior of $\gamma$ as a function of $\delta$. 
 
 From this figure one can notice how the value of the characteristic exponent $\gamma$ is modified as the ratio $g_{\downarrow\uparrow}/g_{\uparrow \uparrow}$ is varied. Here it is important to stress that $\gamma$ is also influenced by the presence of the harmonic confinement. The value of this exponent is reduced as $\delta$ grows, and thus the relaxation process becomes slower in time. Quite remarkable is the fact that for strong disorder strengths $\delta$, the $\gamma$ parameters take very similar values, almost independent of the ratio of the inter and intra-species interaction coupling. This can indicate that the disorder has become the dominant contribution during the elapsed evolution. The inset in Fig. \ref{Figure5} shows the error of the magnetization fit.

\section{Final remarks}
\label{section5}

We have studied the time dynamics of initially localized ferromagnetic domains evolving under the influence of both, disordered confinement and contact interactions. The purpose of such an investigation was to establish the persistence of ferromagnetic order in the domains, namely spatial regions with definite magnetization, when the competition of structural disorder and interactions could lead the system, evolving under their inner dynamics, to nullify such an initial magnetic pattern. To study such a magnetization relaxation process as a function of time, we proposed a model system simulating a double ferromagnetic domain evolving under static disorder. The model consisted of a two-species $^{87}$Rb Bose-Einstein condensate, whose components labeled as $\uparrow$ and $\downarrow$ states, were placed spatially separated, lying each one in the halves of a 2D potential resulting from the superposition of a harmonic potential and a square lattice. The description of the dynamics was addressed within the mean field Gross-Pitaevskii approach, by solving the coupled equations associated to different hyperfine components. To have a reliable analysis of the evolution in time of the magnetization under the presence of disorder, our analysis consisted of an extensive set of numerical calculations over different realizations of non-correlated disorder having a given amplitude $\delta$, and constant values of the intra and inter-species interactions $g_{\downarrow\uparrow}$ and $g_{\uparrow \uparrow}$ respectively. Regarding the magnitude of the intra and inter-species interaction $g_{\downarrow\uparrow}$ and $g_{\uparrow \uparrow}$, we worked in the regime in which the ratio between these coefficients $g_{\downarrow\uparrow}/g_{\uparrow \uparrow}$ guaranty miscibility of hyperfine components, and also considering appropriate coupling interaction strengths away from the strong interaction effects.

Our main conclusion is that the relaxation process of a double ferromagnetic domain, that is loss of magnetization in definite regions of space, becomes slower and slower as the structural disorder is increased, while in contrast, increasing the ratio between inter and intra particle contact interactions  $g_{\downarrow \uparrow}/g_{\uparrow \uparrow}$ tends to degrade the initial state. We reach these result from a robust study of the time evolution of the right and left magnetization of a quantum system described above. Textures or local magnetization, as referred in current literature, are suitable observables to track the effects of the disorder media in systems having more that one component and, also, are accessible physical quantities with single spin resolution techniques \cite{Weitenberg, Boll} used in current experimental setups. 

The manuscript here presented sets a platform for the design of specific protocols appropriate to study demagnetization processes or frustration effects associated to geometry and energy disorder \cite{Pierce,Windsor,Korzhovska,Reinld}. Also, our study aims for the investigation of the dynamics induced by measurement in the sense that sources of disorder can be either internal as those here considered, or external as those associated to reservoirs in contact with assessable quantum systems \cite{Hilary}. We expect that our work will trigger further theoretical analysis as for instance, the long range character proper of the dipole-dipole magnetic interactions, as well as the homogeneous environment where the elemental constituents move. Those aspects still remain as open questions to be addressed. Understanding the dynamics of magnetic domains, have become nowadays a relevant topic not only within the context of the fundamental physics, but also associated to the emergence of technological uses. Practical applications of the investigation here presented are directly related with the design of magnetic logic and memory devices.

\acknowledgements{
This work was partially funded by grant IN10620 DGAPA (UNAM). G.A.D.C., would like to thank V. Romero-Roch\'{\i}n for computational resources at Skyrmion IF-UNAM, and R. Zamora-Zamora for useful guidance. G.A.D.C., L.A.G.G. and C.M., acknowledge scholarship from CONACYT.
}


\begin{thebibliography}{99}
\bibitem{Calculations} The grid used in our numerical calculations was 512 $\times$ 512. All the calculations were implemented in state of the art GPU hardware.
\end{thebibliography}
\end{document}